\documentstyle[12pt]{article}
\def\be {\begin{equation}}
\def\ee {\end{equation}}
\def\ba {\begin{eqnarray}}
\def\ea {\end{eqnarray}}

\def\r  {\rho}

\def\la {\label}
\def\le {\left}
\def\ri {\right}

\def\f {\frac}

\def\bi {\begin{itemize}}
\def\ei {\end{itemize}}
\begin{document}
\def\bea{\begin{eqnarray}}
\def\eea{\end{eqnarray}}
\title{\bf {Logarithmic correction to the Brane equation in Topological
 Reissner-Nordstr\"om de Sitter Space }}
 \author{M.R. Setare  \footnote{E-mail: rezakord@ipm.ir}
  \\{Physics Dept. Inst. for Studies in Theo. Physics and
Mathematics(IPM)}\\
{P. O. Box 19395-5531, Tehran, IRAN }}

\maketitle
\begin{abstract}
In this paper we study braneworld cosmology when the bulk space is
a charged black hole in de Sitter space (Topological
 Reissner-Nordstr\"om de Sitter Space) in general dimension,
then we compute leading order correction to the Friedmann equation
that arise from logarithmic corrections to the entropy in the
holographic-branworld cosmological framwork. Finally we consider
the holographic entropy bounds in this senario, we show the
entropy bounds are also modified by logarithmic term.
 \end{abstract}
\newpage

 \section{Introduction}
Holography is believed to be one of the fundamental principles of
the true quantum theory of gravity\cite{{HOL},{RAP}}. An
explicitly calculable example of holography is the much--studied
AdS/CFT correspondence \cite{AdS}. Unfortunately, it seems that we
live in a universe with a positive cosmological constant which
will look like de Sitter space--time in the far future. Therefore,
we should try to understand quantum gravity or string theory in de
Sitter space preferably in a holographic way. Of course, physics
in de Sitter space is interesting even without its connection to
the real world; de Sitter entropy and temperature have always been
mysterious aspects of quantum gravity\cite{GH}.\\
While string theory successfully has addressed the problem of
entropy for black holes, dS entropy remains a mystery. One reason
is that the finite entropy seems to suggest that the Hilbert space
of quantum gravity for asymptotically de Sitter space is finite
dimensional, \cite{{Banks:2000fe},{Witten:2001kn}}.
 Another, related, reason is that the horizon and entropy in
de Sitter space have an obvious observer dependence. For a black
hole in flat space (or even in AdS) we can take the point of view
of an outside observer who can assign a unique entropy to the
black hole. The problem of what an observer venturing inside the
black hole experiences, is much more tricky and has not been given
a satisfactory answer within string theory. While the idea of
black hole complementarity provides useful clues, \cite
{Susskind}, rigorous calculations are still limited to the
perspective of the outside observer. In de Sitter space there is
no way to escape the problem of the observer dependent entropy.
This contributes to the difficulty of de Sitter space.\\
More recently, it has been proposed that defined in a manner
analogous to the AdS/CFT correspondence,  quantum gravity in a de
Sitter (dS) space is dual to a certain
 Euclidean  CFT living on a spacelike boundary of the
dS space~\cite{Strom} (see also earlier works
\cite{Hull}-\cite{ohta}). Following the proposal, some
investigations on the dS space have been carried out
recently~\cite{Mazu}-\cite{set3}. According to the dS/CFT
correspondence, it might be expected that as the case of AdS black
holes~\cite{Witten2}, the thermodynamics of cosmological horizon
in asymptotically dS spaces can be identified with that of a
certain Euclidean CFT residing on a spacelike boundary of the
asymptotically dS spaces. \\
   There has been much recent interest in calculating the quantum
corrections to $S_{BH}$ (the Bekenestein-Hawking entropy)
\cite{maj1x}-\cite{medved}. The leading-order correction is
proportional to $\ln{S_{BH}}$. There are, {\it two} distinct and
separable sources for this logarithmic correction
 \cite{gg2x,maj3x} (see also recent paper by Gour and Medved \cite{ medved}).
  Firstly, there should be a correction
 to the number of microstates that is a quantum correction to the
 microcanonical entropy, secondly, as any black hole will typically exchange heat or
 matter with its surrounding, there should also be a correction due to thermal
 fluctuations in the horizon area.\\
 In this paper we consider the brane universe in the bulk background of the
 topological Reissner-Nordstr\"om de Sitter (TRNdS) black holes.
 In fact there is pressing cosmological motivation for introducing the CFT potential dual to
 the charge of the black hole. It is, in particular, the presence
 of a non-vanishing charge that can induce the desirable feature
 of a non-vanishing bounce \cite {medv1}.
 At first we find the thermodynamical quantities of the dual CFT,
 then we show that the Friedmann brane equation can be in the
 Cardy-Verlinde formula when the brane crosses the black hole
 horizon or the cosmological horizon. Taking into account thermal
 fluctuations defines the logarithmic corrections to both
 cosmological and black hole horizon entropies. As a result the
 Cardy-Verlinde formula and Friedmann brane equation receive the
 logarithmic corrections.

Therefore here we generalize the logarithmic corrections (with
respect to the temperature) that appear in the five dimensional
case of Ref \cite{od, od1} to any dimension, but in the specific
case of a topological Reissner-Nordstrom black hole in de Sitter
(dS) space. The arguments are based on dS/CFT correspondence for
the counting of microstates, which of course is not a well
established result. However, one may accept this conjecture and
study its consequences. Finally we consider the implication of the
prior analysis with regard to the holographic entropy bounds, we
show the entropy bounds are also modified by logarithmic term.

\section{FRW equation in the background of TRNdS Black Holes}
The topological Reissner-Nordstr\"om dS black hole solution in
$(n+2)$-dimensions has the following form
\begin{eqnarray}
&& ds^2 = -f(r) dt^2 +f(r)^{-1}dr^2 +r^2 \gamma_{ij}dx^{i}dx^{j}, \nonumber \\
&&~~~~~~ f(r)=k -\frac{\omega_n M}{r^{n-1}} +\frac{n \omega_n^2
Q^2}{8(n-1) r^{2n-2}}
     -\frac{r^2}{l^2},
\end{eqnarray}
where
\begin{equation}
\omega_n=\frac{16\pi G}{n\mbox {Vol}(\Sigma)},\hspace{1cm}\phi
=-\frac{n}{4(n-1)}\frac{\omega_n Q}{r^{n-1}},
\end{equation}
where $Q$ is the electric/magnetic charge of Maxwell field, $M$ is
assumed to be a positive constant, $l$ is the curvature radius of
de Sitter space, $\gamma_{ij}dx^idx^j$ denotes the line element of
an $n$-dimensional hypersurface $\Sigma_k$ with the constant
curvature $n(n-1)k$ and its volume $V(\Sigma_k)$. $\Sigma_k$ is
given by spherical ($k=1$), flat ($k=0$), hyperbolic  $(k=-1)$,
$\phi$ is the electrostatic potential related to the charge $Q$.
When $k=1$, the metric Eq.(1) is just the Reissner-Nordstr\"om-de
Sitter solution. For general $M$ and $Q$, the equation $f(r)=0$
may have four real roots. Three of them are real, the largest one
is the cosmological horizon $r_{c}$, the smallest is the inner
(Cauchy) horizon of black hole, the middle one is the outer
horizon $r_{+}$ of the black hole. And the fourth is negative and
has no physical meaning. The case $M=Q=0$ reduces
to the de Sitter space with a cosmological horizon $r_{c}=l$.\\
When $k=0$ or $k=-1$, there is only one positive real root of
$f(r)$, and this locates the position of cosmological horizon
$r_{c}$.\\
In the case of $k=0$, $\gamma_{ij}dx^{i}dx^{j}$ is an
$n-$dimensional Ricci flat hypersurface, when $M=Q=0$ the solution
Eq.(1) goes to pure de Sitter space
\begin{equation}
ds^{2}=\frac{r^{2}}{l^{2}}dt^{2}-\frac{l^{2}}{r^{2}}dr^{2}+r^{2}dx_{n}^{2},
\end{equation}
in which $r$ becomes a timelike coordinate.\\
When $Q=0$, and $M\rightarrow -M$ the metric Eq.(1)is the TdS
(Topological de Sitter) solution \cite{med}, which have
a cosmological horizon and a naked singularity.\\
 For the purpose of getting the Friedmann-Robertson-Walker(FRW) metric, we impose
the following condition\cite{EV},
\begin{equation}
{1\over{f(r)}}\left({{dr}\over{d\tau}}\right)^2-f(r)\left({{dt}\over
{d\tau}}\right)^2=-1, \label{condition2}
\end{equation}
which leads to a timelike brane.
 Substituting Eq.(4) into
the  TRNdS solution  Eq.(1), one has the induced brane metric
which takes FRW form
\begin{equation}
ds^2=-d\tau^2+r^2(\tau)\gamma_{ij}dx^idx^j,
\end{equation}
Timelike brane, i.e a brane that has a Minkowskian metric, can
only cross the black hole horizon. On the contrary, a spacelike
brane, i.e. a brane with Euclidean metric, is able to cross both
the black hole horizon and the cosmological horizon. In order to
derive the $4$-dimensional spacelike brane, the imposed condition
(\ref{condition2}) has to be slightly changed by replacing the
$`-$' with a $`+$' on the right-hand side of it.\\
 The equation of
motion of the brane is given by\cite{SV}
 \begin{equation}
 {\cal K}_{ij}=\frac{\sigma}{n}h_{ij},
  \end{equation}
where ${\cal K}_{ij}$ is the extrinsic curvature, and $h_{ij}$ is
the induced metric on the brane, $\sigma$ is  the brane tension.
The extrinsic curvature, ${\cal K}_{ij}$, of the brane can be
calculated and expressed in term of function $r(\tau)$ and
$t(\tau)$. Thus one rewrites the equations of motion (6) as
\begin{equation}
{{dt}\over{d\tau}}=\frac{\sigma r}{f(r)}.
\end{equation}
 Using Eqs.(4,7), we can drive FRW equation with $H=\frac{\dot{r}}{r}$,
\begin{equation}\label{frweq}
H^2=\frac{f(r)}{r^2}+ \sigma^2={{-\omega_{n}M}\over
r^{n+1}}+{{nw^2_{n}Q^2}\over{8(n-1)r^{2n}}} +{k\over
r^2}-\frac{1}{l^{2}}+\sigma^2,
\end{equation}
where, $H$ is the Hubble parameter. We choose the brane tension
$\sigma=\frac{1}{l}$ to obtain a critical brane. Therefore
Eq.(\ref{frweq} leads to
\begin{equation}\label{frweq1}
H^2={{-\omega_{n}M}\over
r^{n+1}}+{{nw^2_{n}Q^2}\over{8(n-1)r^{2n}}} +{k\over r^2}.
\end{equation}
 Making use of the fact that the metric
for the boundary CFT can be determined only up to a conformal
factor, we rescale the boundary metric for the CFT to be of the
following form
\begin{equation}
ds^2_{CFT}=\lim_{r\to\infty}\left[{l^2\over r^2}ds^2_{n+2}\right]
=dt^2+l^2\gamma_{ij}dx^idx^j.
\end{equation}
Evidently, the Euclidean CFT time must be scaled by a factor
$l/r$. Processing on this basis, the thermodynamic relations
between the boundary CFT and the bulk TRNdS are given by
\begin{eqnarray}
E_{CFT}&=&M{l\over r},\ \ \ \ \ \ \ \ \ \ \Phi_{CFT}=\Phi{l\over
r},
 \cr T_{CFT}&=&{T_{TRNdS}}{l\over r},\ \ \ \ \ \ \
 S_{CFT}=S_{TRNdS},
\end{eqnarray}
where black hole horizon Hawking temperature $T_{TRNdS}^{b}$ and
entropy $S_{TRNdS}^{b}$ are given by
\begin{eqnarray}\label{blacentr}
 && T_{TRNdS}^{b}=\frac{f'(r_{+})}{4\pi} =\frac{1}{4\pi r_+} \left((n-1) -(n+1)\frac{r_+^2}{l^2}
   -\frac{n\omega_n^2 Q^2}{8 r_+^{2n-2}}\right), \nonumber \\
&& S_{TRNdS}^{b} =\frac{r_+^n\mbox{Vol}(\Sigma)}{4G},
\end{eqnarray}
where $r=r_{+}$ is black hole horizon and
$V_{+}=r_{+}^{n}Vol(\Sigma)$ is area of it in $(n+2)-$dimensional
asymptotically dS space.\\Here we review the BBM prescription
\cite{BBM} for computing the conserved quantities of
asymptotically de Sitter spacetimes briefly. In a theory of
gravity, mass is a measure of how much a metric deviates near
infinity from its natural vacuum behavior; i.e, mass measures the
warping of space. Inspired by the analogous reasoning in AdS
space \cite{{by},{b}} one can construct a divergence-free
Euclidean quasilocal stress tensor in de Sitter space by the
response of the action to variation of the boundary metric we have
\begin{eqnarray}
T^{\mu \nu} &=& {2 \over \sqrt{h}} { \delta I \over \delta h_{\mu
\nu}} = \ \  {1 \over 8\pi G} \left[ K^{\mu\nu} - K \, h^{\mu\nu}
+ {n \over l} \, h^{\mu\nu} +\frac{l}{n}  \,G^{\mu\nu} \right] ,
 \label{stressminus}
\end{eqnarray}
where $h^{\mu\nu}$ is the metric induced on surfaces of fixed
time, $K_{\mu\nu}$, $K$ are respectively extrinsic curvature and
its trace, $G^{\mu\nu}$ is the Einstein tensor of the boundary
geometry. To compute the mass and other conserved quantities, one
can write the metric $h^{\mu\nu}$ in the following form
\begin{equation}
    h_{\mu\nu} \, dx^{\mu} \, dx^{\nu } =
       N_{\rho}^{2} \, d\rho^{2} +
       \sigma_{ab}\, (d\phi^a + N_\Sigma^a \, d\rho) \,
               (d\phi^b + N_\Sigma^b \, d\rho)
       \label{boundmet}
\end{equation}
where the $\phi^{a}$ are angular variables parametrizing closed
surfaces around the origin. When there is a Killing vector field
$\xi^{\mu}$ on the boundary, then the conserved charge associated
to $\xi^{\mu}$ can be written as \cite{{by},{b}}
\begin{equation}
   Q =  \oint_{\Sigma}  d^{n}\phi \,\sqrt{\sigma } \,
   n^{\mu}\xi^{\mu} \,T_{\mu\nu}
   \label{chargedef}
\end{equation}
where $n^{\mu}$ is the unit normal vector on the boundary,
$\sigma$ is the determinant of the metric $\sigma_{ab}$. Therefore
the mass of an asymptotically de Sitter space is as
\begin{equation}
    M =
    \oint_{\Sigma}  d^{n}\phi \,\sqrt{ \sigma } \, N_{\rho} \,
\epsilon
    ~~~~~;~~~~~ \epsilon \equiv
    n^{\mu}n^{\nu} \,
    T_{\mu\nu} \, ,
    \label{massdef}
\end{equation}
where Killing vector normalized as $\xi^{\mu} = N_{\rho} n^{\mu}$.
Using this prescription \cite{BBM}, the gravitational mass,
subtracted the anomalous Casimir energy, of the TRNdS solution is
\begin{equation}
\label{3eq3} E^{c}=-M =-\frac{r_c^{n-1}}{\omega_n} \left (k
-\frac{r_c^2}{l^2} +
    \frac{n\omega_n^2 Q^2}{8(n-1)r_c^{2n-2}}\right).
\end{equation}
 The Hawking temperature $T_{TRNdS}^{c}$
and entropy $S_{TRNdS}^{c}$ associated with the cosmological
horizon are
\begin{eqnarray}\label{cosentr}
 && T_{TRNdS}^{c}=\frac{-f'(r_{c})}{4\pi} =\frac{1}{4\pi r_c} \left(-(n-1)k +(n+1)\frac{r_c^2}{l^2}
    +\frac{n\omega_n^2 Q^2}{8 r_c^{2n-2}}\right), \nonumber \\
&& S _{TRNdS}^{c}=\frac{r_c^n\mbox{Vol}(\Sigma)}{4G},
\end{eqnarray}
where $V_{c}=r_{c}^{n}Vol(\Sigma)$ is area of the cosmological
horizon.  The AD mass of TRNdS solution can be expressed in terms
of black hole horizon radius $r_+$ and charge $Q$,
\begin{equation}\label{admass}
 E^{b} =M =\frac{r_+^{n-1}}{\omega_n} \left
(1-\frac{r_+^2}{l^2} +
   \frac{n\omega_n^2 Q^2}{8(n-1)r_+^{2n-2}}\right).
\end{equation}
In terms of the energy density $\rho_{CFT}=E_{CFT}/V$, the
pressure $p_{CFT}=\rho_{CFT}/n$, the charge density
$\rho_{QCFT}=Q/V$ and the electrostatic potential
$\Phi_{CFT}=\Phi{l\over r}$ of the CFT within the volume
$V=r^n{\rm Vol}(\Sigma)$, also the specific heat of the black hole
is given by
 \begin{equation}
 C^{c,b}=\frac{dE^{c,b}}{dT}=\frac{4\pi r_{c,+}^{2}(8 k (1-n)l^{2}r_{c,+}^{n-2}+8 (n+1)r_{c,+}^{n}+
 n\omega_{n}^{2}l^{2}r_{c,+}^{-n}Q^{2}) }
 { \omega_{n}( 8l^{2}(n-1)k+8
 r_{c,+}^{2}(n+1)+(1-2n)l^{2}\omega_{n}^{2}r_{c,+}^{2-2n}Q^{2})}.
 \end{equation}
 As one can see the above specific heat is positive in the case $k=-1,
 k=0$, for $k=1$, $C^{c,b}$ is positive only with following condition
 \begin{equation}
8 (n+1)r_{c,+}^{n}+n\omega_{n}^{2}l^{2}r_{c,+}^{-n}Q^{2} > 8 k
(1-n)l^{2}r_{c,+}^{n-2}
 \end{equation}

The first Friedmann equation take the following form
\begin{equation}\label{firsteq}
H^2={{16\pi G}\over{n(1-n)}}\left(\rho_{CFT}-{1\over
2}\Phi\rho_{QCFT}\right) +{k\over r^2},
\end{equation}

\section{Logarithmic correction to the Cardy-Verlinde formula and FRW brane cosmology in TRNdS
bulk} There has been much recent interest in calculating the
quantum corrections to $S_{BH}$ (the Bekenestein-Hawking entropy)
\cite{maj1x}-\cite{myuag}. The corrected formula takes the form
\begin{equation}
\label{entro} {\cal S}=S_0-\frac{1}{2}{\rm ln }{C}+\ldots
\end{equation}
When $r_{c,+}^{2}\gg l^{2}$, $C\simeq n S_0$, in this case we have
\begin{equation}\label{entro1}
 {\cal S}=S_0-\frac{1}{2}{\rm ln }{S_0}+\ldots
\end{equation}
\\
It is now possible to drive the corresponding correction to
Cardy-Verlinde formula. The Casimir energy $E_C$, defined as
\begin{equation}\label{caseq}
 E_{C}^{c,b}=(n+1) E^{c,b}-nT^{c,b}S^{c,b}-n\phi^{c,b} Q,
 \end{equation}
 in this case, is found to be
 \begin{equation}
 E_{C}^{c,b} =\frac{-2n k r_{c,+}^{n-1}Vol(\Sigma)}{16\pi
G},
\end{equation}
which is valid for both cosmological and black hole horizon. One
can see that the entropy Eqs. (\ref{cosentr},\ref{blacentr}) of
the cosmological and black hole horizon can be written as
\begin{equation}
 S^{c,b}=\frac{2\pi l}{n}\sqrt{|\frac{E_{C}^{c,b}}{k}|(2(E^{c,b}-E_{q}^{c,b})-E_{C}^{c,b})},
\end{equation}
where
\begin{equation}
E_{q}^{c,b} = \frac{1}{2}\phi^{c,b} Q
=-\frac{n}{8(n-1)}\frac{\omega_n Q^2}{r_{c,+}^{n-1}}.
\end{equation}
 For the present discussion, the total entropy is assumed to be of
 the form Eq.(\ref{entro1}), where the uncorrected entropy, $S_{0}$
 correspondence to that associated in Eqs. (\ref{cosentr},\ref{blacentr}). It then follows by
 employing Eqs.(\ref{blacentr}-\ref{admass}) that the Casimir energy Eq.(\ref{caseq}) can
 be expressed in term of the uncorrected entropy. (Following
 expressions are valid for both cosmological and black hole horizon, then for simplicity
 we omit the subscript $c$ and $b$ )
 \begin{equation}
  E_{C} =\frac{-2n r_{c,+}^{n-1}\mbox{Vol}(\Sigma)}{16\pi
G}+\frac{nT}{2}Ln S_{0},
\end{equation}
After some calculation, the total entropy Eq.(\ref{entro1}) to
first order in the logarithmic term, is given by \cite{set5}
  \bea \label{corent}
 S &\simeq&\frac{2\pi
l}{n}\sqrt{|\frac{E_C}{k}|(2(E-E_q)-E_C)}+ \frac{E_{q}[(3n+1)E-2n
E_{q}+(1-2n)E_C]+E[nE_C-(n+1)E]}{4E_C(E-E_q-E_{C}/2)} \nonumber\\
&& Ln \left (\frac{2\pi l}{n}\sqrt{|\frac{E_C}{k}|(2(E-E_q)-E_C)}
\right) \eea Therefore taking into account thermal fluctuations
defines the logarithmic corrections to both cosmological and black
hole entropies. As a result the Cardy-Verlinde formula receive
logarithmic corrections in our interest  TRNdS black hole
background in any dimension.\\
The first Friedmann equation (\ref{firsteq}) can be rewritten in
terms thermodynamical formulas of the CFT on the brane when brane
crosses the cosmological or event horizon \cite{{SV},{set1}}, at
these times the first Friedmann equation coincides with the
Cardy-Verlind formula. As a direct consequence of the logarithmic
correction arising in Eq.(\ref{corent}) the Friedmann equation,
also receives the logarithmic correction due to thermal
fluctuations of the bulk gravity system. The Hubble parameter $H$
is related with the Hubble entropy as
\begin{equation} \label{hubb}
S_H\equiv (n-1){{HV}\over{4G}},
\end{equation}
which is equal with bulk black hole entropy at the moment when the
brane crosses the black hole horizon $r=r_+$ in the case $k=1$,
and crosses the cosmological horizon $r=r_{c}$ for the cases $k=0$
and $k=-1$ \cite{{SV},{set1}}. By substituting Eq.(\ref{corent})
into Eq.(\ref{hubb}) one finds the modified Friedmann equation at
the holographic points \bea H^{2}&=&\frac{16
G^{2}}{(n-1)^{2}V^{2}}S^{2}= \frac{16
G^{2}}{(n-1)^{2}V^{2}}[\left(\frac{2\pi
l}{n}\sqrt{|\frac{E_C}{k}|(2(E-E_q)-E_C)} \right)^{2}+ \frac{4\pi
l}{n}\sqrt{|\frac{E_C}{k}|(2(E-E_q)-E_C)} \nonumber
\\ && \frac{E_{q}[(3n+1)E-2n
E_{q}+(1-2n)E_C]+E[nE_C-(n+1)E]}{4E_C(E-E_q-E_{C}/2)}\nonumber
\\ && Ln\left(
\frac{2\pi l}{n}\sqrt{|\frac{E_C}{k}|(2(E-E_q)-E_C)} \right)\ ],
 \eea
after setting $E_{q}=0$, $n=3$,$k=1$ the above equation is agree
with result of Ref.\cite{od1} for Friedmann brane equation in
5-dimensional Schwarzschild de Sitter bulk which is as following
\bea \label{lnln} H^2 = \left({2 G \over V }\right)^2 \left[
\left( {4\pi l \over 3 \sqrt{2}} \right)^2 \left| E_C\left( E -
{1\over 2} E_C\right)\right| \right. -{4\pi l \over 3 \sqrt{2}} {E
\left( 4E -3E_C \right) \over \left( 2E -E_C \right) E_C } &&
\nonumber
 \\ \left.\sqrt{\left| E_C\left( E
 - {1\over 2} E_C\right)\right|}
\ln \left({4\pi l \over 3 \sqrt{2} }\sqrt{\left| E_C\left( E
 - {1\over 2} E_C\right)\right|}\right)\ \right].
 \eea At the holographic points $r=r_{c,+}$, after setting
$\sigma=\frac{1}{l}$, we have
\begin{equation}\label{uncor}
H^2={1\over l^2}\ \ \ \ \ \ \ {\rm at}\ \ \ \ \ \ r=r_{+ ,  c}.
\end{equation}
Formally, the Friedmann equation.(32) holds precisely at the
instant when the brane crosses black hole and cosmological
horizons. Here we extend the analysis to consider an arbitrary
scale factor $r$ where the world-volume of the brane is given by
the line-element (5). Thus, around each horizons we assume the
Friedmann equation as follows:
 \bea \label{modieq}
H^2&=&{k\over r^2}+{{16\pi
G}\over{n(1-n)}}\left(\rho_{CFT}-{1\over 2}\Phi\rho_{QCFT}\right)
+\frac{32 \pi
G^{2}l}{(n-1)^{2}nV^{2}}\sqrt{|\frac{E_C}{k}|(2(E-E_q)-E_C)}\nonumber\\&&
 \frac{E_{q}[(3n+1)E-2n
E_{q}+(1-2n)E_C]+E[nE_C-(n+1)E]}{4E_C(E-E_q-E_{C}/2)}\nonumber
\\ && Ln \left(
\frac{2\pi l}{n}\sqrt{|\frac{E_C}{k}|(2(E-E_q)-E_C)}\right), \eea
then the logarithmic corrections for the FRW equation are given
by the last term on the right--hand side in terms of the
uncorrected entropy Eqs.(\ref{blacentr},\ref{cosentr}) of the
black hole. Here the logarithmic corrections have been included
up to first-order in the logarithmic term. Therefore, at least,
the brane receives the thermal radiation from the black hole, the
thermal correction should change the dynamics of the brane from
the leading order or zero temperature behavior.
 \\
Verlinde pointed out that the FRW equation (\ref{frweq1}) can be
related to three cosmological entropy bounds
\begin{equation}\label{bheq}
S_{BH}=(n-1)\frac{V}{4GR},\hspace{1cm} \mbox{Bekenstein-Hawking
bound}
\end{equation}

\begin{equation}\label{bveq}
S_{BV}=-\frac{2\pi R}{n}(E-\frac{2\pi
G_{n}Q^{2}}{(n-1)V}),\hspace{1cm} \mbox{Bekenstein-Verlinde bound}
\end{equation}
and Hubbel bound which is given by Eq.(\ref{hubb}). Here $G_{n}$
is the gravitational constant in bulk which is given by
\begin{equation}\label{greq}
G_{n}=\frac{Gl}{n-1}.
\end{equation}
The FRW equation (\ref{frweq1}) can be rewritten as
\begin{equation}\label{cvfrw}
S_{H}=\sqrt{S_{BH}(S_{BH}-2S_{BV})},
\end{equation}
similarly we can rewrite the modified Friedmann Eq.(\ref{modieq})
as
\begin{equation}\label{cvfrw1}
S_{H}=\sqrt{S_{BH}(kS_{BH}-2S_{BV})+A S_{c}  \ln (S_c)},
\end{equation}
where
\begin{equation}\label{cvent}
S_{c}=\frac{2\pi l}{n}\sqrt{|\frac{E_{c}}{k}|(2(E-Eq)-E_c)},
\end{equation}
\begin{equation}\label{coeq}
A=\frac{E_{q}[(3n+1)E-2n
E_{q}+(1-2n)E_c]+E[nE_c-(n+1)E]}{4E_c(E-E_q-E_{c}/2)}.
\end{equation}
Now if we consider $k=1$, also if we conjecture the redefined
Bekenstein-Hawking entropy as
\begin{equation}\label{entmodi}
S_{BH}\rightarrow S'_{BH}=S_{BH}-\frac{A S_c \ln
S_c}{2(S_{BV}-S_{BH})},
\end{equation}
in this case Eq.(\ref{cvfrw1}) can be rewritten as following
\begin{equation}\label{cvfrw2}
S_{H}=\sqrt{S'_{BH}(S'_{BH}-2S_{BV})},
\end{equation}
therefore the entropy bounds are also modified by logarithmic
term.

  \section{Conclusion}
  One of the striking results for the dynamic dS/CFT correspondence
is that the Cardy-Verlinde's formula on the CFT-side coincides
with the Friedmann equation in cosmology when the brane crosses
the cosmological or event horizon $r=r_{c,+}$ of the topological
Reissner-Nordstr\"om black hole. This means that the Friedmann
equation knows the thermodynamics of the CFT (Since conformal
symmetry in the bulk is broken by the presence
  of a black hole, a prospectively dual boundary theory is, strictly speaking, not
  necessarily a conformal one. Nonetheless, for convenience sake, we continue to refer to
  the relevant boundary theories as CFT). There is pressing cosmological
  motivation for introducing the CFT potential dual to
 the charge of the black hole. Such models are of significant interest because they allow
 for the possibility of a non-singular bounce (as opposed to a big bang/ crunch)
 \cite{medv1,pelo}.  \\
 For a large class of black hole, the Bekenstein-Hawking entropy
 formula receives additive logarithmic corrections due to thermal
 fluctuations. On the basis of general thermodynamic arguments, Das et al
 \cite{das2x} deduced that the black hole entropy can be expressed as
\be {\cal S} = \ln \r = S_0 - \f{1}{2} \ln \le( C~T^2 \ri) +
\cdots \la{corr3}. \ee
  In this paper we have analyzed this correction of the entropy of TRNdS
 black hole in any dimension in the light of dS/CFT. We have obtain the logarithmic
 correction to both cosmological and black hole entropies. Then using the form of the
 logarithmic correction Eq.(\ref{entro1}) one can show the corresponding correction to the Cardy-Verlinde formula which relates the
 entropy of a certain CFT to its total energy and Casimir energy in arbitrary dimension.
As a direct consequence of the logarithmic correction arising in
Eq.(\ref{corent}) the Friedmann equation, also receives the
logarithmic correction due to thermal fluctuations of the bulk
gravity system. Moreover, we have consider the holographic entropy
bounds in this scenario, we have shown that the entropy bounds are
also modified by logarithmic term.\\
It should be mentioned that in standard cosmology where there are
no corrections, the first term in right hand side of equation.(35)
represent the curvature contribution to the brane motion. The
second term can be regarded as the contribution from the
radiation and it redshifts as $r^{-4}$ for a brane moving in the
$5-$dimensional TRNdS bulk background. The last term in the
right-hand side of equation.(35), goes like $r^{-6}$, it is
dominant at early times of the brane evolution while at late
times the second term, i.e. the radiative matter term, dominates
and thus the last term can be neglected. At this point a couple
of questions are raised. First, how the additional term in the
Hubble equation.(35) which come from thermal fluctuations, change
the dynamics of the brane? The second question arises when one
includes both semiclassical (the self-gravitational
effect\cite{setvag}) and logarithmic corrections. Which is
dominant correction and when does this dominant take place during
the brane evolution? We hope to address these interesting issues
in a future work.

\end{document}